%% file: ms.tex
\documentclass[12pt,preprint]{aastex}
\shorttitle{Spectral index of high-energy cosmic rays}
\shortauthors{Amenomori et al.}

\begin{document}

\title{New estimation of the spectral index of high-energy cosmic rays\\ 
as determined by the Compton-Getting anisotropy}

\author{
M.~Amenomori,$^{1}$ X.~J.~Bi,$^{2}$ D.~Chen,$^{3}$ S.~W.~Cui,$^{4}$
Danzengluobu,$^{5}$ L.~K.~Ding,$^{2}$ X.~H.~Ding,$^{5}$ C.~Fan,$^{6}$
C.~F.~Feng,$^{6}$ Zhaoyang Feng,$^{2}$ Z.~Y.~Feng,$^{7}$
X.~Y.~Gao,$^{8}$ Q.~X.~Geng,$^{8}$ H.~W.~Guo,$^{5}$ H.~H.~He,$^{2}$
M.~He,$^{6}$ K.~Hibino,$^{9}$ N.~Hotta,$^{10}$ Haibing~Hu,$^{5}$
H.~B.~Hu,$^{2}$ J.~Huang,$^{11}$ Q.~Huang,$^{7}$ H.~Y.~Jia,$^{7}$
F.~Kajino,$^{12}$ K.~Kasahara,$^{13}$ Y.~Katayose,$^{3}$
C.~Kato,$^{14}$ K.~Kawata,$^{11}$ Labaciren,$^{5}$ G.~M.~Le,$^{15}$
A.~F.~Li,$^{6}$ J.~Y.~Li,$^{6}$ Y.-Q.~Lou,$^{16}$ H.~Lu,$^{2}$
S.~L.~Lu,$^{2}$ X.~R.~Meng,$^{5}$ K.~Mizutani,$^{13,17}$ J.~Mu,$^{8}$
K.~Munakata,$^{14}$ A.~Nagai,$^{18}$ H.~Nanjo,$^{1}$
M.~Nishizawa,$^{19}$ M.~Ohnishi,$^{11}$ I.~Ohta,$^{20}$
H.~Onuma,$^{17}$ T.~Ouchi,$^{9}$ S.~Ozawa,$^{11}$ J.~R.~Ren,$^{2}$
T.~Saito,$^{21}$ T.~Y.~Saito,$^{22}$ M.~Sakata,$^{12}$
T.~K.~Sako,$^{11}$ M.~Shibata,$^{3}$ A.~Shiomi,$^{9,11}$
T.~Shirai,$^{9}$ H.~Sugimoto,$^{23}$ M.~Takita,$^{11}$
Y.~H.~Tan,$^{2}$ N.~Tateyama,$^{9}$ S.~Torii,$^{13}$
H.~Tsuchiya,$^{24}$ S.~Udo,$^{11}$ B.~Wang,$^{8}$ H.~Wang,$^{2}$
X.~Wang,$^{11}$ Y.~Wang,$^{2}$ Y.~G.~Wang,$^{6}$ H.~R.~Wu,$^{2}$
L.~Xue,$^{6}$ Y.~Yamamoto,$^{12}$ C.~T.~Yan,$^{11}$ X.~C.~Yang,$^{8}$
S.~Yasue,$^{25}$ Z.~H.~Ye,$^{15}$ G.~C.~Yu,$^{7}$ A.~F.~Yuan,$^{5}$
T.~Yuda,$^{9}$ H.~M.~Zhang,$^{2}$ J.~L.~Zhang,$^{2}$
N.~J.~Zhang,$^{6}$ X.~Y.~Zhang,$^{6}$ Y.~Zhang,$^{2}$ Yi~Zhang,$^{2}$
Zhaxisangzhu,$^{5}$ and X.~X.~Zhou$^{7}$ \\
(The Tibet AS$\gamma$ Collaboration)}

\affil{
$^{1}$Department of Physics, Hirosaki University, Hirosaki 036-8561, Japan \\
$^{2}$Key Laboratory of Particle Astrophysics, Institute of High Energy Physics, Chinese Academy of Sciences, Beijing 100049, China \\
$^{3}$Faculty of Engineering, Yokohama National University, Yokohama 240-8501, Japan \\
$^{4}$Department of Physics, Hebei Normal University, Shijiazhuang 050016, China \\
$^{5}$Department of Mathematics and Physics, Tibet University, Lhasa 850000, China \\
$^{6}$Department of Physics, Shandong University, Jinan 250100, China \\
$^{7}$Institute of Modern Physics, SouthWest Jiaotong University, Chengdu 610031, China \\
$^{8}$Department of Physics, Yunnan University, Kunming 650091, China \\
$^{9}$Faculty of Engineering, Kanagawa University, Yokohama 221-8686, Japan \\
$^{10}$Faculty of Education, Utsunomiya University, Utsunomiya 321-8505, Japan \\
$^{11}$Institute for Cosmic Ray Research, University of Tokyo, Kashiwa 277-8582, Japan \\
$^{12}$Department of Physics, Konan University, Kobe 658-8501, Japan \\
$^{13}$Research Institute for Science and Engineering, Waseda University, Tokyo 169-8555, Japan \\
$^{14}$Department of Physics, Shinshu University, Matsumoto 390-8621, Japan \\
$^{15}$Center of Space Science and Application Research, Chinese Academy of Sciences, Beijing 100080, China \\
$^{16}$Physics Department and Tsinghua Center for Astrophysics, Tsinghua University, Beijing 100084, China \\
$^{17}$Department of Physics, Saitama University, Saitama 338-8570, Japan \\
$^{18}$Advanced Media Network Center, Utsunomiya University, Utsunomiya 321-8585, Japan \\
$^{19}$National Institute of Informatics, Tokyo 101-8430, Japan \\
$^{20}$Tochigi Study Center, University of the Air, Utsunomiya 321-0943, Japan \\
$^{21}$Tokyo Metropolitan College of Industrial Technology, Tokyo 116-8523, Japan \\
$^{22}$Max-Planck-Institut f\"ur Physik, M\"unchen D-80805, Deutschland \\
$^{23}$Shonan Institute of Technology, Fujisawa 251-8511, Japan \\
$^{24}$RIKEN, Wako 351-0198, Japan \\
$^{25}$School of General Education, Shinshu University, Matsumoto 390-8621, Japan
}

\begin{abstract}
The amplitude of the Compton-Getting (CG) anisotropy contains 
the power-law index of the cosmic-ray energy spectrum. Based 
on this relation and using the Tibet air-shower array data, 
we measure the cosmic-ray spectral index to be $-3.03 \pm 0.55_{stat} \pm
 < 0.62_{syst}$ between 
6 TeV and 40 TeV, consistent with $-$2.7 from direct energy 
spectrum measurements. Potentially, this CG anisotropy 
analysis can be utilized to confirm the astrophysical origin 
of the ``knee'' against models for non-standard hadronic 
interactions in the atmosphere.
\end{abstract}


\section{Introduction} \label{intro}
The cosmic-ray (CR) energy spectrum has been observed using ground-based 
air shower 
detectors and various instruments on board satellites and balloons.
It features a power law d$N$/d$E$ $\propto$ $E^{- \gamma}$ in the
energy range from 30 GeV to 100 EeV, where
$N$ denotes the number of cosmic rays at the top of the atmosphere, $E$ the 
cosmic-ray particle energy.
The spectral index $\gamma$ 
changes from 2.7 to 3.1 around the ``knee'' at $E \simeq$ 4 PeV.
Various models have been proposed for the origin of this knee;
most models suppose that it is of astrophysical origin \cite{Hoerandel2},
while there are some models that attribute it to non-standard
high-energy nucleon interactions in the Earth's atmosphere.

For a power-law CR energy spectrum, the Compton-Getting 
(CG) anisotropy involves the spectral index $\gamma$.
When a detector observing cosmic rays moves relative to the rest frame 
of the CR plasma in the Earth orbit around the Sun,
a fractional CR intensity variation $\Delta I$ can be observed in the solar time frame. 
For a presumed power-law CR energy spectrum, this CG anisotropy \cite{CG-1,CG-2} is 
expressed as
\begin{equation}
\frac{\Delta I}{<I>} = (\gamma + 2)\frac{v}{c}\cos\theta, \label{eq-1}
\end{equation}
where $I$ denotes the CR intensity, 
$\gamma$ the power-law index of the cosmic-ray energy spectrum,
$v$ the orbital velocity, $c$ the speed of light and 
$\theta$ the angle between the arrival direction of CRs and
the moving direction of the observer.

At multi-TeV energies, a clear evidence 
for the CG anisotropy in the solar diurnal variation was first reported in \cite{C-G}.
The observed variation was in reasonable agreement with a sinusoidal curve 
expected from the CG anisotropy with the maximum phase of the curve deviating 
from 6:00 h by $+$2 hours at 2$\sigma$ significance. This
deviation was ascribed to meteorological effects on the underground muon 
intensity.
The Tibet air-shower experiment performed the first study on energy dependence of the solar 
diurnal variation and showed that it was consistent with the CG anisotropy 
above multi-TeV energies \cite{UDO,COROTATE}. 

Conventionally, the spectral index above $\sim$100 TeV is obtained by indirect measurements with air-shower
arrays, in which each CR particle energy is
determined from its shower size. The relation between particle energy and
shower size is 
deduced from air-shower simulations assuming hadronic interaction models 
as a reasonable
extrapolation of the accelerator results to the higher energy range. 
The spectral index thus obtained is valid only when no violation of the 
standard hadronic interaction is expected in the relevant energy range.
This fact leaves a room for an argument whether the knee is due to non-standard 
high-energy nucleon interactions in the atmosphere or not.

On the other hand, the CG anisotropy does not depend on CR particle energy by Eq.(\ref{eq-1}). 
We can then measure the power-law index of the all-particle 
energy spectrum in the way it is virtually uninfluenced by hadronic 
interaction models, because air-shower simulations are used only for the purpose of 
estimating the rough energy range where we are observing the CG anisotropy.
Potentially, this method makes it possible to confirm that the bend of the 
energy spectrum at the knee is of astrophysical origin, not due to non-standard hadronic 
interactions in the atmosphere.

\section{Tibet Air-Shower Experiment}
The Tibet air-shower experiment has been operating successfully 
at 90.522$^\circ$ E, 30.102$^\circ$ N and 4300 m above sea level since 1990.
Being upgraded several times, the Tibet III array we use here
was completed in November 1999 \cite{as1,as4}. 
It consists of 533 scintillation counters of 0.5 m$^2$ each. 
Each counter is viewed by a fast-timing (FT) photomultiplier tube and is placed 
on a 7.5 m square grid with an enclosed area of 22,050 m$^2$ \cite{as5}.
A 0.5 cm thick lead plate is put on top of each counter to improve fast-timing
data by converting gamma rays into electron-positron pairs.
An event trigger signal is issued when any fourfold coincidence takes place in the
FT counters recording more than 0.6 particle, which results in the trigger rate of 
about 680 Hz at a few-TeV threshold energy.
We estimate the energy of a primary CR particle by $\sum \rho_{\mbox{\scriptsize FT}}$, 
which is the sum of the number of particles/m$^2$ for each FT counter. 

\section{Cosmic-Ray Data Analysis}
We collected $7.7 \times 10^{10}$ events during 1319 live days from November
1999 to November 2005. After some data selections (i.e., software trigger
condition of 
any fourfold coincidence in the FT counters recording more than 0.8 particle 
in charge, zenith angle of the arrival direction less than 45$^{\circ}$, 
and air shower core position located in the array), 
the events left were histogrammed in hourly bins in the local solar time 
according to their event time and incident direction.
To suppress the seasonal change in the daily variation, we corrected the 
number of events for each month
considering monthly variations of the live time.

The daily and yearly event rates vary by $\pm$ 2\% and $\pm$ 5\% \cite{as5}
respectively, affected mostly by meteorological effects. To remove these
temporal variations, we adopted the following East$-$West subtraction procedure \cite{EW}.
We first obtain $E(t)$ and $W(t)$, the daily variation at solar time frame for each of East 
and West incident events, according to the geographical longitude 
of incident direction of each event. Subsequently dividing $E(t) - W(t)$ by 2$\delta t$, 
an hour-angle separation between the mean directions of E- and W-incident events, we finally 
reach $D(t)$, the differential of the physical variation at solar time frame $R(t)$.
\begin{equation}
D(t) \equiv \frac{E(t) - W(t)}{2\delta t} = \frac{R(t+\delta t) - R(t-\delta t)}{2\delta t}
     = \frac{\mbox{d}}{\mbox{d}t} R(t)  \label{EWeq}
\end{equation}
The advantage of this method is that meteorological effects and 
possible detector biases which are expected to produce common variations 
for both E- and W-incident events largely cancel out.

\section{Results and Discussion}
The CR events were divided into eight 
$\sum \rho_{\mbox{\scriptsize FT}}$ bins in order to find out the energy 
region 
where the observed solar daily variation is free from effects other than 
the CG anisotropy.
The fitting curve for the observed solar daily variation is expressed as
\begin{equation}
f(\lambda) = \alpha \cos\left(\frac{\pi}{12}(\lambda - \phi) \right), \label{eq-2}
\end{equation}
where $\alpha$ denotes the amplitude of the CG anisotropy,
$\lambda$ [hr] the local solar time and  
$\phi$ [hr] the phase where the sinusoidal curve reaches its maximum.
Note that in a differential form, the phase is shifted earlier 
by 6 hours (1/4 cycle) from the corresponding actual daily variation and 
the amplitude is $\pi$/12 times that of the physical CG anisotropy. 

\begin{figure}
  \begin{center}
    \includegraphics[width=8.2cm]{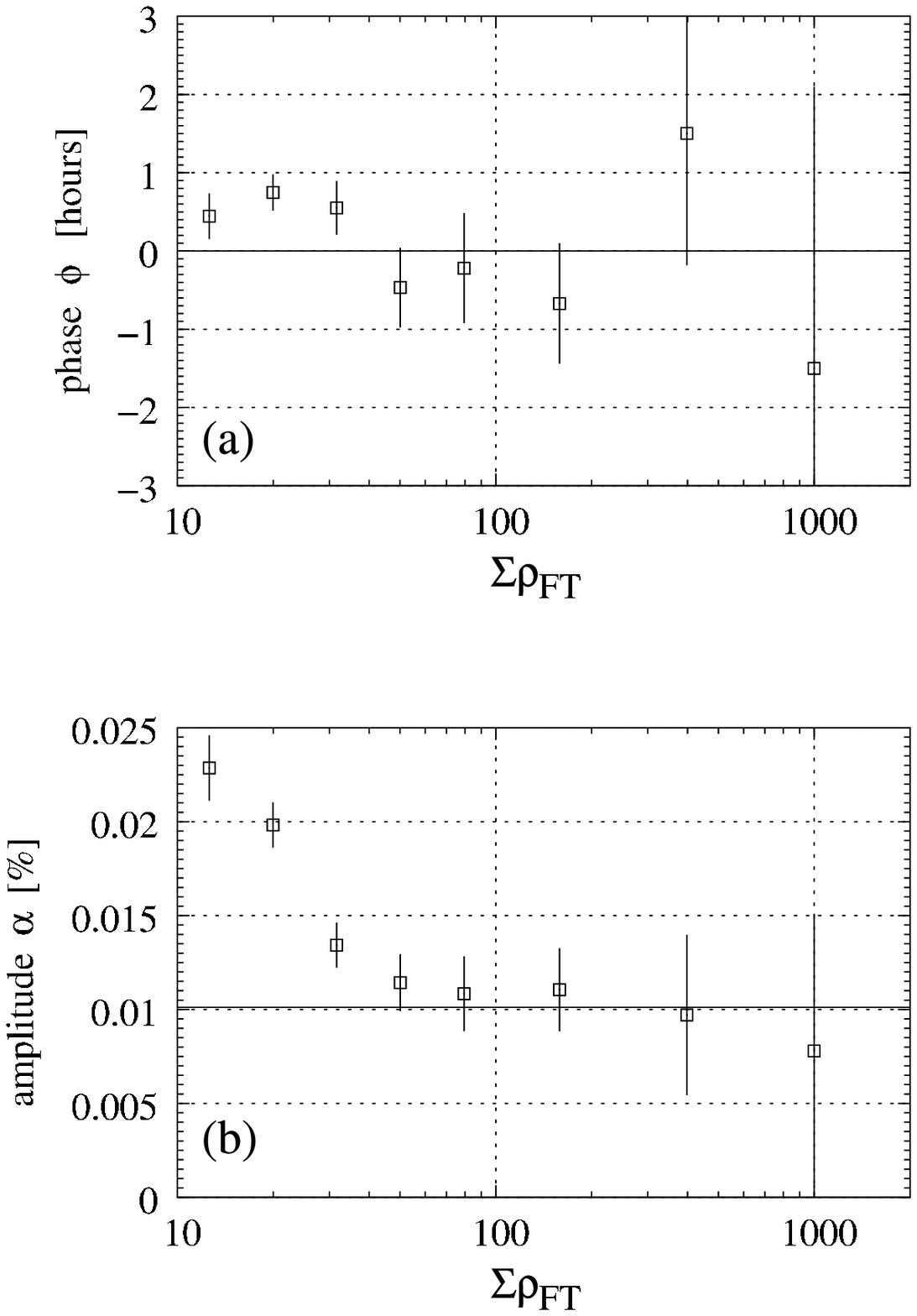}
    \caption{\label{fig1}(a) : $\sum \rho_{\mbox{\scriptsize FT}}$ dependence
      of the phase $\phi$, (b) : $\sum \rho_{\mbox{\scriptsize FT}}$ dependence 
      of the amplitude $\alpha$. Both are evaluated in a differential form. The 
      solid lines show the expected values from the CG anisotropy, assuming 
      $\gamma = -$2.7. The error bars are statistical only.}
  \end{center}
\end{figure}

Figure \ref{fig1}(a) shows $\sum \rho_{\mbox{\scriptsize FT}}$ dependence
 of the phase $\phi$ in Eq.(\ref{eq-2}).
Although there exists deviation due to some contamination in the region of
$\sum \rho_{\mbox{\scriptsize FT}} < 50$,
the measured phase is consistent with the expected value of the CG anisotropy in 
$\sum \rho_{\mbox{\scriptsize FT}} \ge 50$.
Therefore, we regard the observed variation in $\sum \rho_{\mbox{\scriptsize FT}} \ge 50$ 
as caused by the CG anisotropy.
Figure \ref{fig1}(b) shows $\sum \rho_{\mbox{\scriptsize FT}}$ dependence
 of the observed amplitude $\alpha$.
Since the CR energy spectrum in the multi-TeV region is expressed 
by a single power law, the amplitude should be consistent with a constant line if there is no
contamination. 
This figure confirms that we do observe the CG anisotropy 
in the region of $\sum \rho_{\mbox{\scriptsize FT}} \ge 50$. 
Thus, we report here on the spectral index using $9.7 \times 10^{9}$ events 
with $\sum \rho_{\mbox{\scriptsize FT}} \ge 50$.  
This region is roughly between 6 TeV and 40 TeV, corresponding to 20\% and 80\% of 
the median energy distribution, respectively.

\begin{figure}
  \begin{center}
    \includegraphics[width=8.2cm]{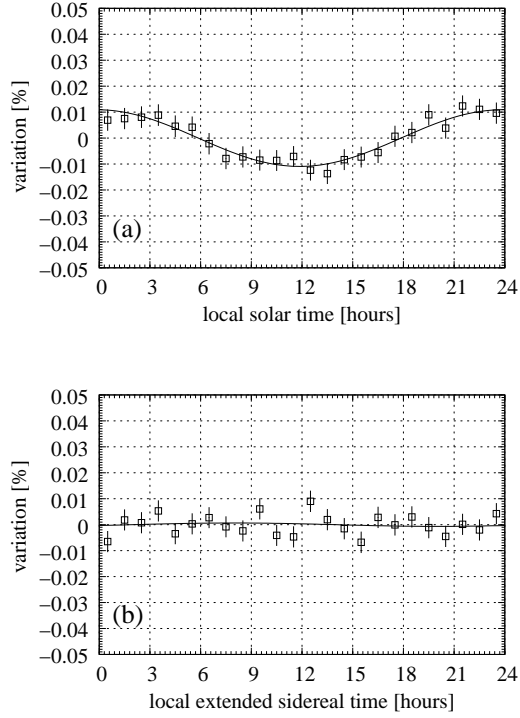}
    \caption{\label{fig2}(a) : The differential variation of relative CR intensity 
      in the local solar time $D(t)$, (b) : the differential variation of relative CR intensity 
      in the local extended-sidereal time.
      The solid lines are the sinusoidal curves best fitted to the data. The error bars 
      are statistical only.}
  \end{center}
\end{figure}

Figure \ref{fig2}(a) shows the observed solar daily variation data fitted
with a sinusoidal curve.
The $\chi^2$-fitting results are summarized in Table I.
From the amplitude of the solar daily variation, 
we obtained $\gamma$ of CRs using the following relation,
\begin{equation} 
\alpha = (\gamma + 2) \frac{v}{c} \frac{\pi}{12} F, \label{eq-3}
\end{equation} 
where $\alpha$ denotes the amplitude, 
$v = 2.978 \times 10^4$ [m/s] the average orbital velocity of the Earth 
and $F = 0.827$ an effective geometrical factor by which $\alpha$ decreases, 
calculated according to the latitude at Yangbajing site (see Appendix).
The factor $\pi$/12 emerges, because we measure the anisotropy in the differential 
form Eq.(\ref{EWeq}).
As a result, we found $\alpha = (10.9 \pm 1.2_{stat}) \times 10^{-5}$.
In Eq.(\ref{eq-3}), we omitted a minor correction.
Strictly speaking, the orbital velocity of the Earth is time dependent.
Although it varies $\pm$2\%, we used the average $v$ instead.
It may become necessary to take this correction into account, 
when more data are accumulated.

A seasonal change of the sidereal daily variation 
due to the galactic anisotropy could produce a spurious variation in the solar time frame. 
The differential variation in the local extended-sidereal time (367.2422 cycles/yr)
is shown in Figure \ref{fig2}(b).
The insignificant variation in the extended-sidereal time $(0.7 \pm 1.2_{stat}) \times 10^{-5}$
ensures that contamination in the solar daily variation 
due to the seasonal change of the sidereal daily variation is less than 12\%
of the CG anisotropy. Adding this as a systematic error, we found
$\gamma = 3.08 \pm 0.55_{stat} \pm < 0.62_{syst}$. 
This systematic error would become smaller in the future when more data are accumulated. 

As the detection efficiency of the Tibet air-shower array depends on CR nuclei 
in this energy range (6 $-$ 40 TeV), the correction for it should be taken into account.
An air-shower simulation employing direct observational data for primary 
cosmic rays revealed that
$-$0.05 should be added to the $\gamma$ above \cite{HD4}. This correction is not needed, 
however, above $\sim$60 TeV where the detection efficiency becomes independent of nuclei, 
reaching $\sim$100\%. 

Finally, the spectral index $\gamma$ between 6 and 40 TeV turned out to be 
$3.03 \pm 0.55_{stat} \pm < 0.62_{syst}$, consistent with 2.7 from direct energy spectrum 
measurements.

Using this CG anisotropy method, a future high-statistics experiment with
a huge effective area 
will be able to measure the power-law index of the all-particle energy spectrum at higher energies.
The spectral index $\gamma$ measured by this method can be compared 
not only with those $\gamma$ values by direct measurements below $\sim$100 TeV 
but also with those by indirect air-shower analyses above $\sim$100 TeV.
This would check the mutual consistency between different ways to measure
the CR energy spectrum.
Following the same strategy, this CG anisotropy analysis can be applied to much higher energies to
confirm the astrophysical origin of the knee.

\acknowledgments
The collaborative experiment of the Tibet Air Shower Arrays has been
performed under the auspices of the Ministry of Science and Technology
of China and the Ministry of Foreign Affairs of Japan. This work was
supported in part by Grants-in-Aid for Scientific Research on Priority
Areas (712) (MEXT), by the Japan Society for the Promotion of Science,
by the National Natural Science Foundation of China, and by the
Chinese Academy of Sciences.


\appendix

\section{APPENDIX}
\indent Coefficient $F$ in Eq.(\ref{eq-3}) is a correction factor for the
latitude at Yangbajing site.
To calculate $F$, 
cos $\theta$ in Eq.(\ref{eq-1}) ($\theta$: the angle between the arrival 
direction of CRs and
the moving direction of the observer) appears as 
\begin{eqnarray}
\cos\theta &=& \cos L \cos \varphi \sin(\varphi + \psi) - \sin T \sin L \
sin \varphi \nonumber \\
           & & - \cos T \cos L \sin \varphi \cos (\varphi + \psi) \label{eq-4} \\
\mbox{with}~~ \psi    &=& 2 \pi t / 24,~~~
\varphi = 2 \pi t / 24 / 365.2422, \nonumber
\end{eqnarray}
where
$L = 30.102^{\circ}$ denotes the latitude at the Tibet air shower array,
$T = 23.44^{\circ}$ the tilt angle of the Earth's spin axis relative to the orbital axis,
$t$ [hr] the lapse time from the beginning of a year,
$\psi$ the solar time and
$\varphi$ a variable which advances from 0 to 2$\pi$ in a year.
Suppose the Earth's spin axis were perpendicular to the ecliptic plane 
($T = 0^{\circ}$) for simplicity,
Eq.(\ref{eq-4}) would reduce to $\cos \theta = \cos L \sin \psi$.
This means the amplitude of CG anisotropy decreases by a factor $F = \cos L = 0.865$.
In reality, however, we need to take into account the tilt of the Earth's spin axis 
according to Eq.(\ref{eq-4}).
We found $F = 0.827$, approximately 4\% smaller than 0.865, by numerically
integrating Eq.(\ref{eq-4}) so that cos $\theta$ can be expressed in terms of the solar time $\psi$ only.

\clearpage

\input{tab1.tex}

\clearpage

\clearpage

\end{document}

%% file: tab1.tex
\begin{table}
  \caption{The $\chi^2$-fitting results of the differential variation in 
    the local solar time (a) and in the local extended-sidereal time (b), assuming 
    the sinusoidal curve Eq.(\ref{eq-2}). The error bars are statistical only.}
  \label{table}
  \begin{center}
    \begin{tabular}{rccc}
      \tableline
      \tableline
      &   $\alpha~(\times~10^{-3}$\%) & $\phi$ [hr] & $\chi^2$/d.o.f. \\
      \tableline
      (a)  &         $10.9 \pm 1.2$ & $-0.14 \pm 0.41$ & 8.03/22 \\
      (b)  &  $0.7 \pm 1.2$ & $-4.0 \pm 6.9$ & 21.9/22 \\
      \tableline
    \end{tabular}
  \end{center}
\end{table}